\documentclass[12pt]{article}
\usepackage{amsmath}
\usepackage[margin=2.0cm]{geometry}
\usepackage{epsfig}
\usepackage{amsfonts}
\usepackage{fancyhdr}
\usepackage{setspace}
\usepackage{enumitem}
\usepackage{graphicx}
\usepackage{subcaption}
\usepackage[nolists, nomarkers]{endfloat}

\usepackage{bm}
\usepackage{cancel}
\usepackage{wasysym}
\usepackage{amssymb}
\usepackage{wrapfig}
\usepackage{textcomp}
\usepackage{arydshln}
\usepackage{tikz}
\usepackage{authblk}
\usepackage{titlesec}
\usepackage{hyperref}
\usepackage{cite}
\usepackage[page,toc,titletoc,title]{appendix}
\usetikzlibrary{arrows}
\newcommand{\authorname}{Vineet Dawara} % NOTE: Remember to change this to your name!
\titleformat{\section}
{\normalfont\bfseries\scshape}{\thesection}{1em}{}

\titleformat{\subsection}
{\normalfont\bfseries\scshape}{\thesubsection}{1em}{}

\titleformat{\subsubsection}
{\normalfont\slshape}{\thesubsubsection}{1em}{}

\titleformat{\title}{\normalfont\bfseries}{\thesection}{1em}{}

\title{\vspace{-3em} \large \bfseries  \textsc{Porosity governs failure in bioconsolidated space bricks}}
\author[1]{\normalsize \authorname}
\author[1]{\normalsize Nitin Gupta}
\author[2]{\normalsize Arjun Dey}
\author[1]{\normalsize Aloke Kumar}
\author[1]{\normalsize Koushik Viswanathan\thanks{Corresponding author: koushik@iisc.ac.in}}
\affil[1]{\normalsize \textsl{Dept. of Mechanical Engineering, Indian Institute of Science, Bangalore}}
\affil[2]{\normalsize \textsl{UR Rao Satellite Centre, Indian Space Research Organization, Bangalore}}
\date{\normalsize \textsc{\today}}

\begin{document}

\maketitle
\thispagestyle{plain}
\hrulefill

\begin{abstract}
  Understanding the mechanical response and failure of consolidated extra-terrestrial soils requires analyses of the interactions between propagating cracks the material's inherent pore structure. In this work, we investigate the fracture behaviour of lunar soil simulant consolidated using microbially induced calcite precipitation (MICP). We develop a numerical framework, based on a lattice network with local beam elements, to simulate the nucleation, propagation, branching and merging of multiple cracks within the sample. Our simulations capture the effects of local pores on crack paths as well as provides a means to predict the behaviour of samples with varying global porosity and/or uncertainities in local material stiffness. We identify multiple statistical lattice parameters that encode signatures of single or multiple crack growth events. Our results reveal the complexities involved in the fracture process with porous brittle solids and may easily be adapted to understand failure mechanisms and micro/macro crack evolution in other consolidated structures.
\end{abstract}

\hrulefill

\section{Introduction}

% (1st para on space exploration and habitat)

The need for exploring extraterrestrial environments as potential candidates for future human habitation has led to significant recent research activity. Most notably, the possibility of \emph{in situ} resource utilization---exploiting locally available materials on extra-terrestrial surfaces---for building human-habitable structures has led to the development of several soil/regolith simulants, especially for the lunar surface \cite{zheng2009cas, mckay1994jsc, weiblen1990preparation, ryu2018development, engelschion2020eac, venugopal2020invention}. The \emph{in situ} use of local regolith as a starting block for building load bearing structures then necessitates the need for developing consolidation and binding mechanisms that are at once robust enough and yet simple enough to implement in an extra-terrestrial environment with minimal human intervention. Several such candidate techniques are being explored, ranging from sintering \cite{farries2021sintered, phuah2020ceramic, taylor2018sintering, hintze2009lunar} and irradiation \cite{kim2021microstructural, altun2021additive} to chemical and biological syntheses \cite{roberts2021blood, roedel2014protein, toutanji2005strength, castelein2021iron}.

% (2nd para on space bricks and their properties)

In our recent work, we have demonstrated the efficacy of a sustainable route for \emph{in situ} consolidation that exploits microbially induced calcite precipitation (MICP) \cite{achal2009strain, ghosh2019sporosarcina}, especially in the context of lunar regolith simulants \cite{dikshit2020microbially, dikshit2021space}. Most notably, the MICP process, when coupled with certain biopolymers, can result in significant compressive strengths ($\sim 5$ MPa) in the final bricks, comparable with that of ice. These bricks themselves---termed \lq space bricks\rq\ and henceforth referred to as such---are quite capable of forming building blocks for complex structures. We have also explored modifications of the MICP route for other simulants, such as for Martian regolith with promising results \cite{dikshit2022microbial}.

% Remember to cite this: gualtieri2015compressive

% (3rd para on need to understand fracture behaviour and methods for the same)

The possibility of using these consolidated bricks, formed via MICP or otherwise, for building more complex, potentially interlinked, structures for actual human habitation requires more comprehensive evaluation of their mechanical properties. While uniaxial strength (compressive in this case) by itself provides a measure of load bearing capacity, it an idealized quantity that is obtained under the assumption that the material is pore-free and completely homogeneous \cite{Knott_1971}. This shortcoming is particularly acute with consolidated bricks, given their inherent porosity \cite{gualtieri2015compressive}. An analogous problem arises when analyzing their failure mechanisms and strength---the pore network plays a central role in determining crack paths and, consequently, effective compressive strengths. These phenomena can seldom be studied experimentally since accurate control of sample porosities is often impossible to effect, so that recourse must be taken to numerical simulations. 

%(4th para on numerical methods)

In this work, we develop a numerical framework for analyzing the mechanical behaviour and failure of consolidated space bricks. Our framework exploits a lattice network model to mimic the effects of both inherent porosity and potential consolidation/bonding statistics. This approach is in stark contrast to more conventional finite element methods that are ill-equipped for dealing with non-uniformly distributed porosities.  While lattice network models can be traced back to humble origins \cite{Hernnikoff}, significant recent developments \cite{alava2006statistical, Arcangelis, HerrmannBond} have enable applications to fracture problems in materials ranging from concrete to bone \cite{Wang, mayya2016splitting}. Consequently, while we compare our results to tests on MICP-consolidated lunar regolith---space bricks---our model framework is applicable to a host of other consolidation schemes as well as more general disordered solids.

This manuscript is organized as follows. We first summarize recent experimental results on space bricks and describe their mechanical properites and failure (Sec.~\ref{sec:background}). The numerical simulation framework is then described in Sec.~\ref{sec:numerical}. The main results pertaining to fracture of space bricks, the accompanying statistics, as well as strength-porosity correlation predictions are presented in Sec.~\ref{sec:results}. We discuss the primary implications of our results as well as the broader applicability of our framework in Sec.~\ref{sec:discussion}.

\section{Experimental: Consolidation and strength of space bricks}
\label{sec:background}

\begin{figure}[h!]
	\centering
	\includegraphics[width=\textwidth]{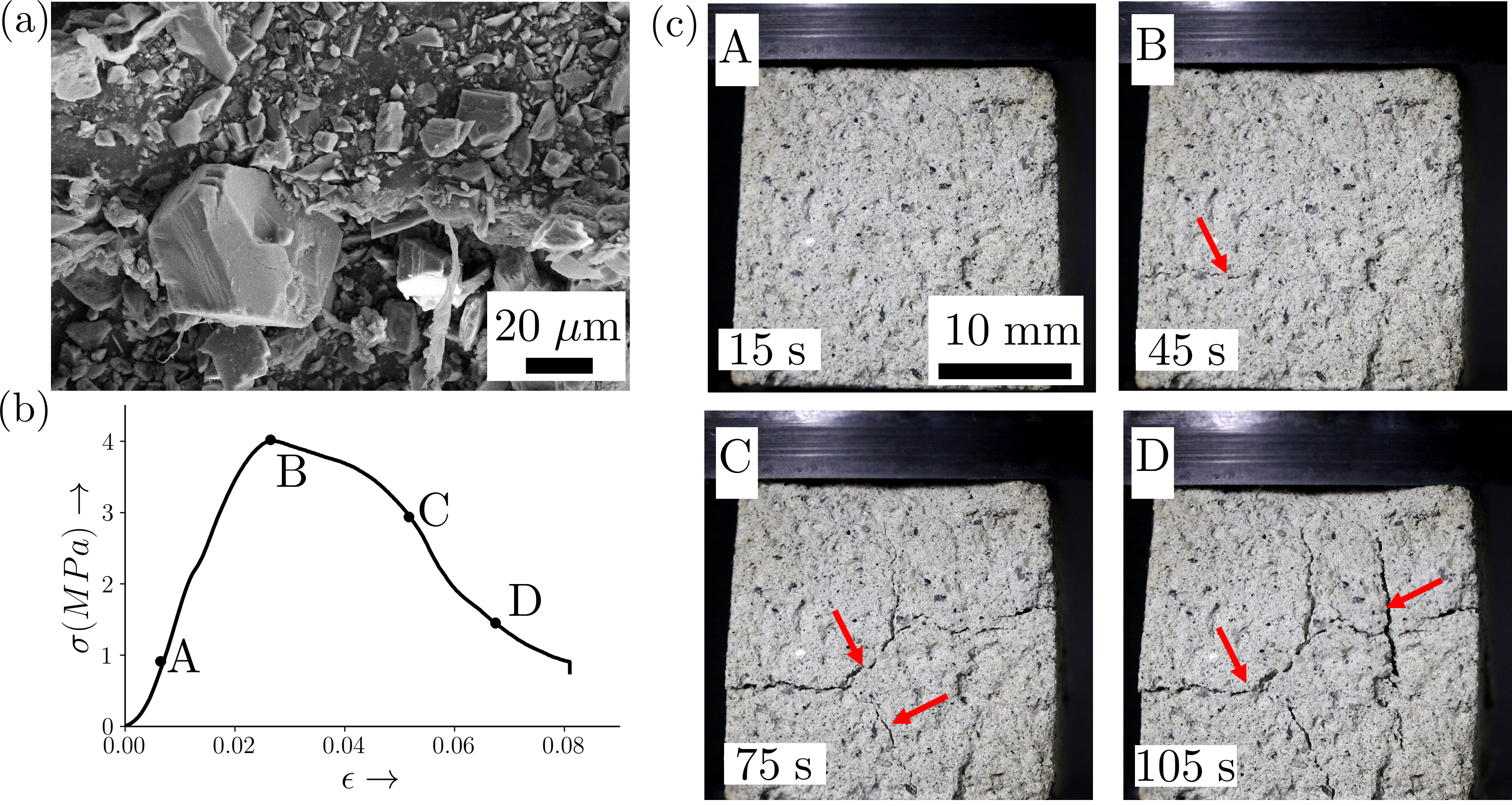}
	\caption{Experimental details of MICP-consolidated space bricks (a) SEM image of (unconsolidated) LSS particles showing polydispersity and large shape variation, (b) typical stress-strain curve for compression loading of a single cubic space brick, and (c) sequence of images showing evolving crack patterns in bricks during compression. Points $A-D$ marked in the frames in (c) correspond to the points marked on the stress-strain curve in panel (b).}
	\label{fig:experimental}
\end{figure}

Lunar regolith is primarily composed of fine-grained basaltic and anorthosite rock particles with traces of minerals like plagioclase and pyroxene. For the development of consolidated space bricks in our recent work as well as in the present study, we use ISRO developed ISAC-1 lunar soil simulant (LSS) as the source material \cite{venugopal2020invention}. A typical SEM image (Carl Zeiss AG-ULTRA 55, Germany) of the as-received LSS sample is shown in Fig.~\ref{fig:experimental}(a). The grains are predominantly polyhedral and polydisperse and chemically comprised of anorthorite. 

The procedure for consolidating the LSS into space bricks involves casting a premixed slurry into a mould, followed by post processing, and has been described in detail elsewhere \cite{dikshit2021space, dikshit2022microbial}. To evaluate the internal porosity and pore size distribution of consolidated samples, independent X-ray micro CT imaging was performed with a smaller sample consolidated under identical conditions (Bruker skyscan 1272). A resolution of 2.5 $\mu$m was used with a window size of 3.2 mm by 4 mm. The specific slurry used for the results summarized in Fig.~\ref{fig:experimental} was comprised of 50g LSS with admixture of 1\% (w/v) guar gum. The result was a single consolidated cubic sample of dimensions 25$\times$25$\times$25 mm, which was then subjected to uniaxial compressive testing (Instron 5697, 5 kN load cell, 1 mm/min).

A typical stress-strain graph during the uniaxial test is shown in Fig.~\ref{fig:experimental}(b). The curve shows a typical stress peak that corresponds to the sample's compressive strength. A sequence of corresponding images are presented in Fig.~\ref{fig:experimental}(c); the instants at which these images were recorded are marked $A-D$ on the stress-strain graph of Fig.~\ref{fig:experimental}(c). It is clear from this sequence that at the peak load, a large primary crack begins to develop (arrow, frame $B$). With subsequent deformation, multiple branches of this crack develop and the sample progressively loses capacity to bear additional load, resulting in the decline between points $B$ and $D$. At $D$, significant crack growth and branching has occurred (see arrows), resulting in complete loss of strength in the sample. This type of multiple crack nucleation and growth events is characteristic of the fracture and failure of brittle solids, wherein a single crack is seldom seen \cite{sammis1986failure}. Similar graphs are obtained for a multiple realizations using the same initial regolith conditions; in all cases, final failure occurred inevitably via the nucleation and propagation of multiple cracks. 

\section{Numerical Framework}
\label{sec:numerical}

In this section, we describe a numerical simulation framework for simulating the fracture and failure behaviour of MICP-consolidated space bricks. The method uses a beam-based multiple degree of freedom lattice model \cite{HerrmannBond} and can be easily extended to study failure in other consolidated simulant systems or materials with porosity. In a suitable limit, our formulation may be considered as a discretization of a micropolar continuum \cite{Bolander}. Additional details of the individual elements can be found in Refs.~\cite{schlangen1997fracture, Hermann}.

\subsection{Beam elements}
\begin{figure}[h!]
	\centering
	\begin{subfigure}[b]{0.55\columnwidth}
		\includegraphics[width=\textwidth]{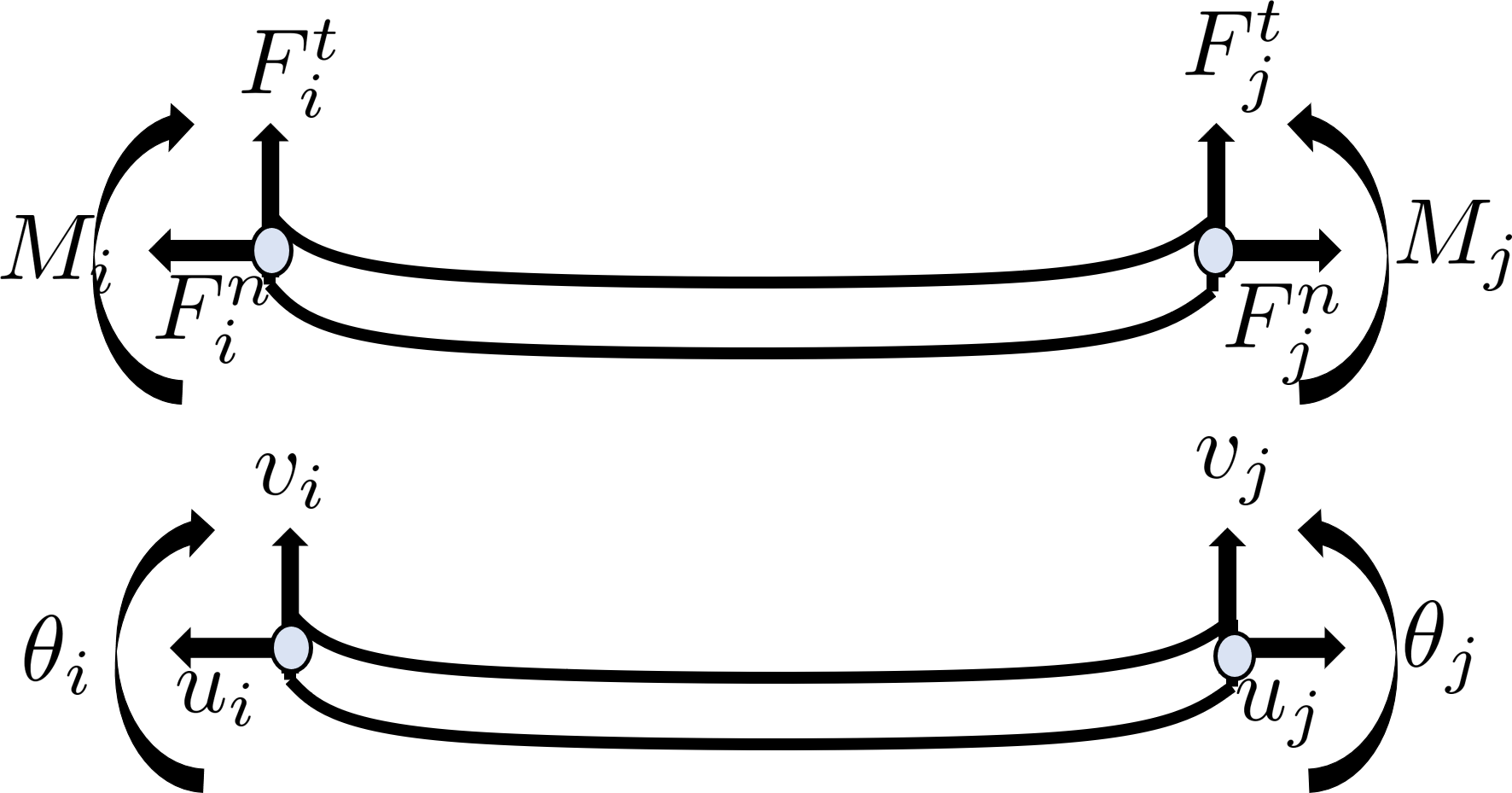}
		\caption{Beam element}
	\end{subfigure}
	\begin{subfigure}[b]{0.4\columnwidth}
		\includegraphics[width=\textwidth]{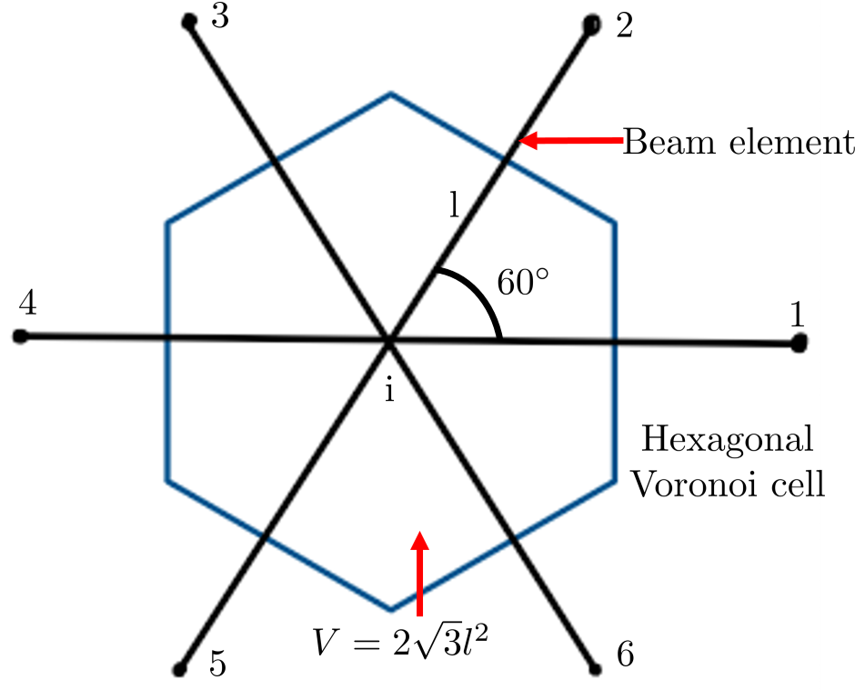}
		\caption{Triangular unit cell}
	\end{subfigure}
	\caption{Details of beams elements and node configuration used for the lattice network.}
	\label{fig:elements}
\end{figure}

A single brick sample is represented as a lattice network, comprising of Euler-Bernoulli beam elements between pre-defined nodes, that can transfer normal, shear, and bending moments, see Fig.~\ref{fig:elements}(a). Nodes are located on a regular triangular lattice with a maximum coordination number of 6, see Fig.~\ref{fig:elements}(b). Nodes represent material while the beams connecting them account for local extensional/rotational degrees of freedom. The potential energy of the beam element connecting nodes $i$ and $j$ in terms of nodal displacements ($u_i, v_i$) and rotation ($\theta$) is as follows:
\begin{align}
\Phi_{ij} = \frac{1}{2}\left[k_n(u_i - u_j)^2 + k_t(v_i - v_j + l\theta_i)^2 + k_tl(\theta_i - \theta_j)(v_i - v_j + l\theta_i) + k_\theta(\theta_i - \theta_j)^2\right] \label{eq: energy}
\end{align}
In the above energy expression, the normal, shear and rotational stiffness are given by
\begin{equation}
k_n = \frac{EA}{l}\quad\quad
k_t = \frac{12EI}{l^3}\quad\quad
k_\theta = \frac{4EI}{l^3}
\end{equation}
where $E$ is Young's modulus; $l$ is length of the beam element; $A$ and $I$ are respectively, the area and moment of inertia of the beam cross-section. The total energy of the lattice network is written by summing the energy in Eq.~\ref{eq: energy} over all elements in the lattice. Consequently, for a complete network, application of external load results in deformation of each of the constituent elements. Equilibrium configuration of the network is hence determined by minimizing the total energy with respect to nodal displacements, resulting in a set of linear algebraic equations of the form
\begin{equation}
  \label{eqn:governing}
\begin{bmatrix}
\bm{K}
\end{bmatrix}_{N\times N}\begin{bmatrix}
\bm{u}
\end{bmatrix}_{N\times 1} = \begin{bmatrix}
\bm{B}
\end{bmatrix}_{N\times 1}
\end{equation} 
Here the boundary conditions are encoded in the vector $\bm{B}$ and the stiffness matrix $\bm{K}$ contains the spring constants for the entire network. In situations of interest, $\bm{K}$ is symmetric and positive definite matrix, so that the Conjugate Gradient Method may be used to invert Eq.~\ref{eqn:governing} efficiently. The result is a specification of the (unknown) generalized displacement vector $\bm{u}$ for the entire system. Note that:
\begin{equation}
 \bm{u}_i \equiv [
u_i, v_i, \theta_i]
\end{equation}
for all nodes $i$.

To this (essentially elastic) system, one must append a fracture criterion to break bonds under specific loading conditions. These then account for local cracks within the system. Since any brittle material, such as our brick, fractures when the local tensile stress exceeds a material-dependent limit, we set a similar condition for each beam in the network: a beam element is removed from the network when the maximum tensile stress ($\sigma_b$) in the beam element exceeds a predefined critical value ($\sigma_c$), mathematically,
\begin{align}
\sigma_b = \frac{F_n}{A} + \alpha \frac{\max|M_i,Mj|}{Z} \geq \sigma_c\label{eq: tensilestress}
\end{align} 
where $F_n$ is the normal force; $Z$ is section modulus of the beam element. The factor $0< \alpha \leq 1$ accounts for the relative importance of bending vis-\'a-vis extension and can be used to simulate the effect of one over the other. For the results reported in this work, we use $\alpha=0.5$, the results reported remain qualitatively unchanged when $\alpha$ is varied over the entire range \cite{schlangen1995experimental}. 

\subsection{Model initialization and mapping with brick properties}
In order to capture the fracture behaviour of our space bricks, we have assumed a uniform 2D triangular beam network with $60^o$ angle between adjacent beams, \emph{cf}. Fig.~\ref{fig:elements}(b). A 2D network is chosen since the experimental loading condition is one of plane strain, and our model boundary conditions are also specified in terms of displacements. The disordered nature of the material can be incorporated either by using uniform spring stiffness on an irregular lattice or using a regular lattice with varying stiffness; our approach uses the latter. Beams are normalized to have length $l = 1$, and square cross-section, with stiffness ($k_n, k_t, k_\theta$) and breaking strength $\sigma_c$. For the results reported here, stiffness ratios are set as $k_t/k_n = 0.1$ and $k_\theta/k_t = 0.33$. 

\begin{figure}[h!]
	\centering
	\includegraphics[width=\linewidth]{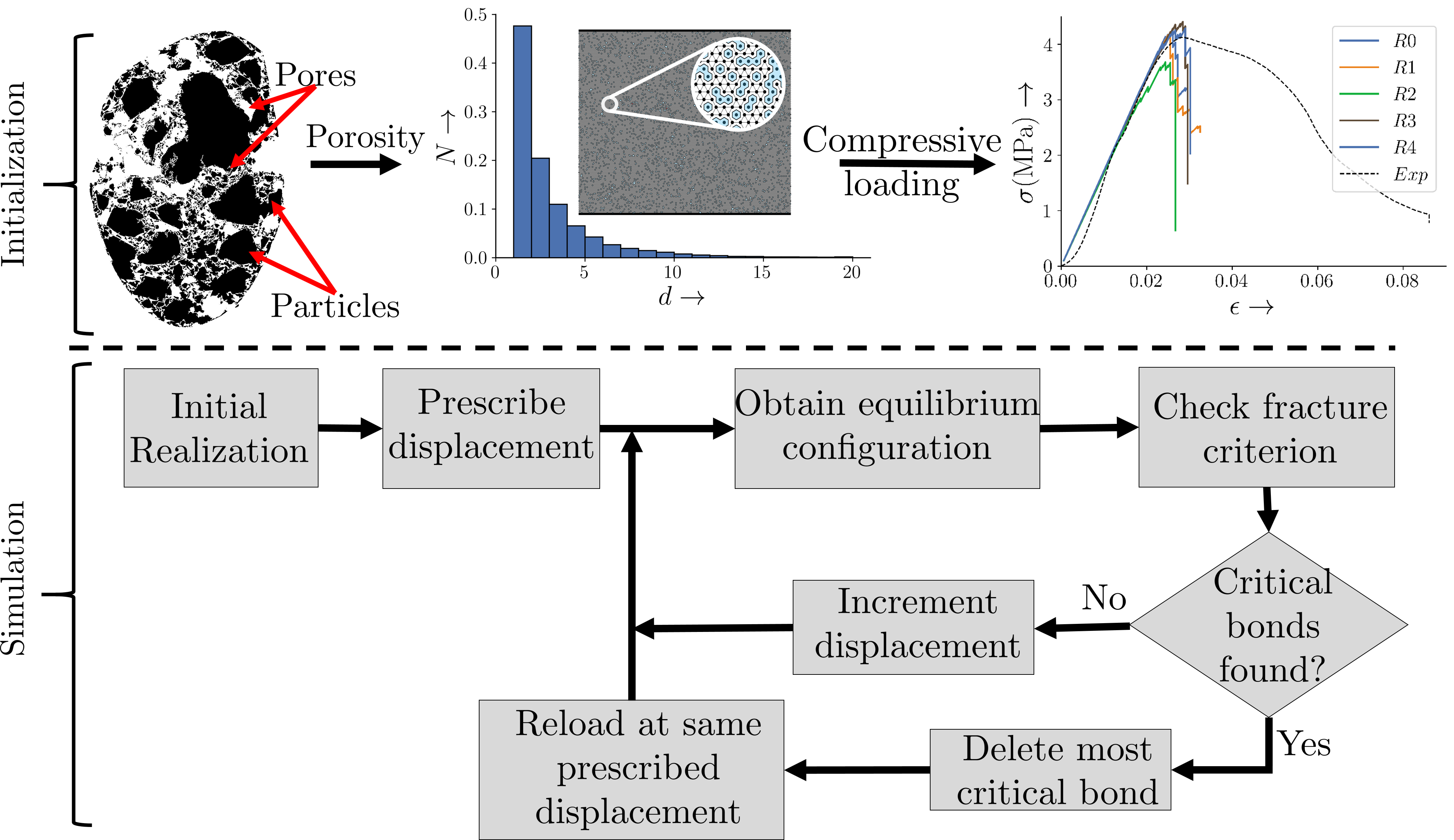}
	\caption{Schematic showing model initiation (top row) and single simulation run (bottom row).}
	\label{fig:modelDetails}
\end{figure}

The pore network within a single brick is generated as follows, see Fig.~\ref{fig:modelDetails}(top row). We first construct a Voronoi tessellation of our uniform triangular lattice, consisting of an array of hexagonal polygons, see Figure ~\ref{fig:elements}(b). Next, pores are introduced by randomly removing nodes and their corresponding Voronoi polygons based on uniform probability $p$, equal to the gross porosity obtained from X-ray micro-CT data described in Sec.~\ref{sec:background}, see Fig.~\ref{fig:modelDetails}(top row, a). The result is a lattice network with a random distribution of pores with global porosity $p$ and a distribution of pore sizes $d$, see Fig.~\ref{fig:modelDetails}(b). Inset to this image shows a single lattice network realization with an enlarged view of the pore network. The pore size $d$ is defined as
\begin{align}
  \label{eqn:poreSize}
d = \frac{\sqrt{3}\times\text{perimeter of pore}}{6l}
\end{align}

Using this scheme, different realizations were generated with global porosity $p=10, 15, 20, 25\%$ and a $300\times 300$ lattice. The average pore size $d$ is summarized in Table\ref{table1} and shows that the mean pore size increases with porosity, as is to be expected.
\begin{table}[h!]
  \caption{Pore size distribution for different global porosities based on 3 lattice realizations.}
  \centering
  \begin{tabular}{|c| c| c|}
    \hline
    $\%$ Porosity & Mean pore size (d$_{mean}$ $\pm$ stddev.)\\
    \hline
    10 & 1.38 $\pm$ 0.36 \\
    15 & 1.67 $\pm$ 0.61 \\
    20 & 2.08 $\pm$ 0.91 \\
    25 & 2.62 $\pm$ 1.36 \\
    \hline
  \end{tabular}
  \label{table1}
\end{table}
 
Uniform compression is simulated using displacement control on the top boundary and vertically constraining nodes on the bottom row. Both the top and bottom nodes are free to move in the horizontal direction. Given the non-dimensionalization of $\bm{K}$ described above, the only material properties that are to be determined are the normal stiffness $k_n$, which controls the slope of stress-strain response, and beam strength $\sigma_c$, which decides the global breaking strength. We adjust these two parameters to match the experimentally obtained stress-strain curve for compression of brick samples, see Fig.~\ref{fig:modelDetails}(top row, c). Five different realizations were used, and the paramters obtained were $p = 30\%, k_n = 125$ and $\sigma_c = 65$.

Once the model is initialized, a single simulation step is as follows, see Fig.~\ref{fig:modelDetails}(bottom row): (a) The initial lattice realization is created, (b) the displacement is prescribed to horizontal boundary nodes, (c) the equilibrium configuration is obtained by solving the matrix equation (\ref{eqn:governing}). Following this, critical beam elements that satisfy the fracture criterion \ref{eq: tensilestress} are determined. (e) if critical beam elements are found, then the maximum stressed one is removed from the network by simply setting its stiffness values to zero, and the matrix $\bm{K}$ is updated. The system is again re-equilibrated at the same prescribed displacement; (f) if no such critical elements are found, the prescribed displacement is increased and $\bm{B}$ is updated. This entire procedure is repeated for every boundary displacement step.

\section{Results}
\label{sec:results}
We now present the results of our numerical investigation of the fracture behaviour of space bricks. It is pertinent to note that the model can, apart from reproducing gross stress-strain curves, also provide significant insight into the statistics of local rupture events as well as crack paths under compressive loading.

\subsection{Crack paths and fracture pattern in space bricks}
\begin{figure}[h!]
\centering
\includegraphics[width=\textwidth]{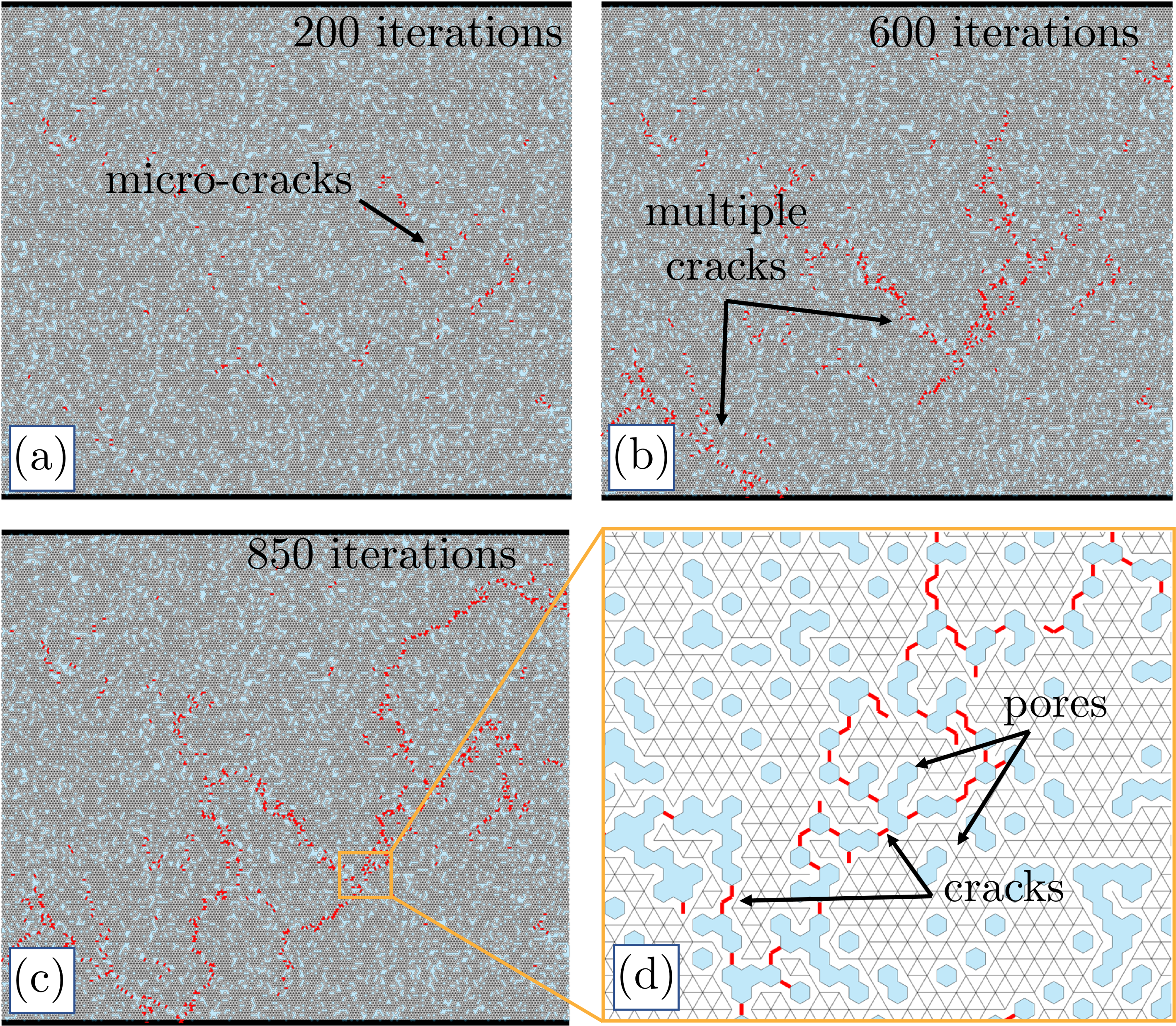}
\caption{Simulated fracture pattern for 25$\%$ porosity with a $200\times 200$ lattice network. Several micro-cracks initiate at or near pore-mediated stress concentrators (panel a) and merge to form multiple macro cracks (panel b). Failure occures when the macro cracks traverse the entire length of the sample (panel c). Inset showes how micro-cracks connect adjacent pores to form a macro-crack.}
\label{fig:fractureSequence}
\end{figure}

Compressive loading of space bricks was simulated using a $200\times 200$ lattice and $p = 20\%$, and a typical sequence of lattice states is shown in Fig.~\ref{fig:fractureSequence}. In the initial stages of the deformation, multiple cracks initiate at different locations within the lattice, depending on the local spatial distribution of pore sizes, shapes, and orientations, see panel (a). These \lq micro-cracks\rq\ are significant because they can result in local stress relaxation without any signature on the global stress-strain curve. As the boundary loading is increased further, additional microcracks emerge within the lattice. However a small section of them begin to form a connected \lq macro-crack\rq\ (panel (b)). These merged cracks grow and tend to follow the loading (vertical) direction, and transition to a horizontal direction. This transition is unlikely what is seen in continuous or mildly porous ($p\leq 5\%$) materials where primary/merged cracks tend to remain aligned with the loading direction \cite{ashby1986failure, sammis1986failure, ashby1990damage}. With increased loading, the larger cracks continue to grow at the expense of smaller micro-cracks until the entire sample is fractured (panel (c)). It is noteworthy to compare the pattern seen in this lattice realization with that in the experimental images, \emph{cf.} Fig.~\ref{fig:experimental}(c). It is interesting that the macro cracks appear to be largely comprised of micro-cracks connecting nearby pores, see inset to panel (c). Clearly, the occurrence of micro and macro cracks appears to be captured (within stochastic variations) by the model. 
      
\subsection{Role of porosity and stiffness variation in determining strength}
\begin{figure}[ht!]
	\centering
	\begin{subfigure}[b]{0.49\columnwidth}
		\includegraphics[width=\linewidth]{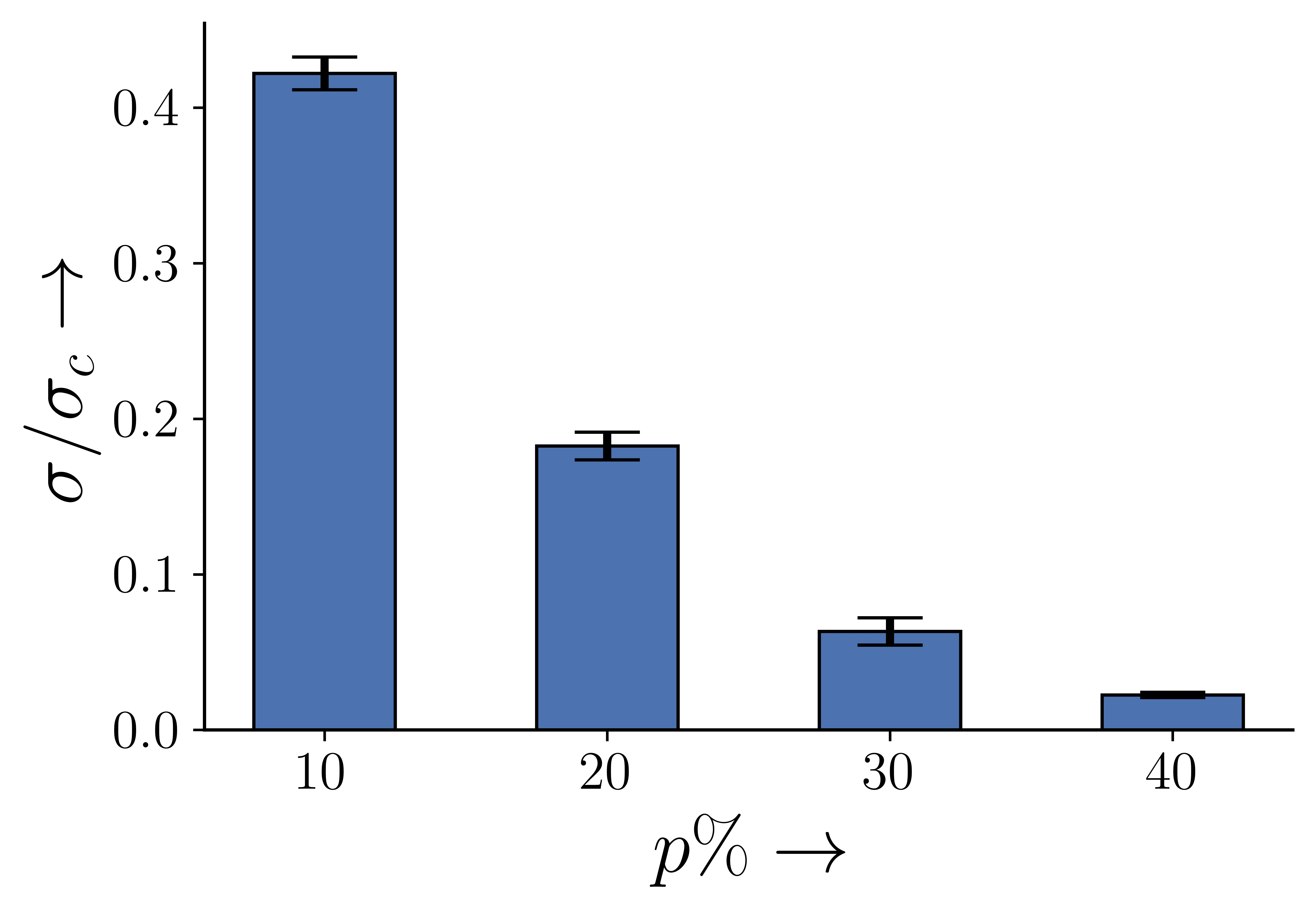}
		\caption{}
	\end{subfigure}
	\begin{subfigure}[b]{0.49\columnwidth}
		\includegraphics[width=\linewidth]{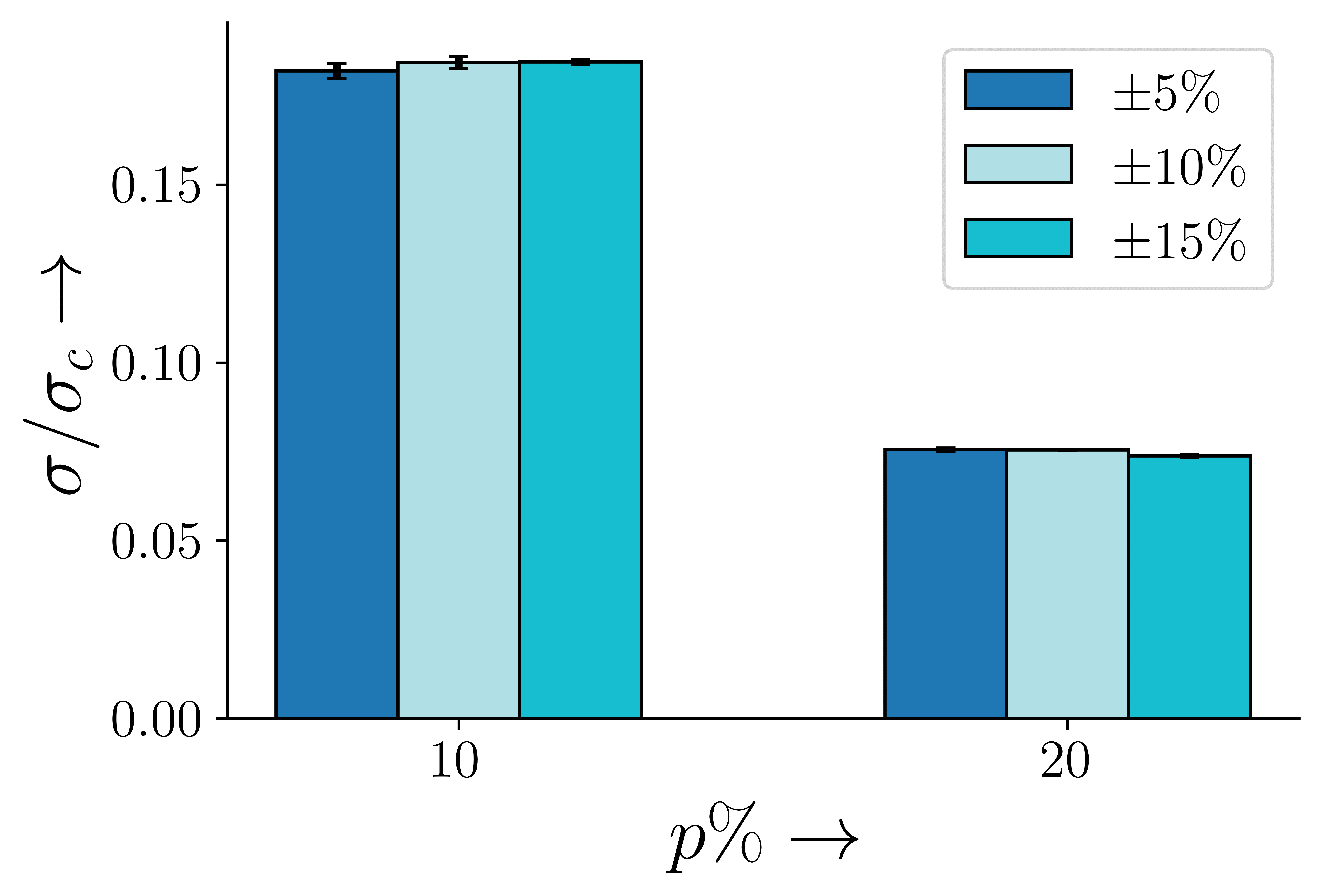}
		\caption{}
	\end{subfigure}
	\caption{(a) Effect of porosity on strength of bulk brick samples as obtained from lattice network simulations. (b) Sample strength as a function of local uncertainities in elastic stiffness. Simulation parameters used: 200$\times$ 200 lattice, $k_n = 125 \pm x\%$ where $x = 5, 10, and 15$, $k_t/k_n = 0.1$, $k_\theta/k_t = 0.33$, $\sigma_b = 65$. Averaged over 3 lattice realizations (varying $k_n$) with the same initial pore structure in each case.}
	\label{fig:porosity}
\end{figure}

It is clear from both the crack propagation results above, as well as the experimental observations, that porosity is expected to play a central role in determining both the strength and failure mechanisms of the bricks. To quantify the role, we simulated the deformation response for various values of the global porosity $p$, ranging from $10\%$ to $40\%$, see Fig.~\ref{fig:porosity}(a). The sample breaking strength $\sigma$ is here represented in non-dimensional form, non-dimensionalized by the individual bond breaking strength $\sigma_c$, see Eq.~\ref{eq: tensilestress}.  For an ideal lattice (material with $p=0$), fracture will occur catastrophically when $\sigma = \sigma_c$. The addition of local pores results in local stress concentrators that can cause premature failure and, consequently, lower $\sigma$. Exactly how much lower is shown in the bar chart of Fig.~\ref{fig:porosity}(a); the error bars correspond to standard deviation values over 3 different lattice realizations. It is clear from this figure that doubling the porosity from $10\%$ to $20\%$ more than halves the compressive strength, with a similar dramatic decrease with subsequent increase in $p$.

Within the specific context of consolidated bricks, we use the lattice simulations to evaluate yet another possibility; that of varying local bond strengths. It is possible during the MICP process that various polymorphs of CaCO$_3$ are generated as binding bridges between individual soil particles. These polymorphs can have varying local elastic stiffness, depending on the nature and orientation of the local bonds. We simulated this uncertainity by varying the normal stiffness $k_n$ within the network by $\pm$5, $\pm$10, $\pm$15\% to the mean stiffness value. For each case, the porosity spatial distribution was kept the same and three different realizations simulated, with uniformly distributed $k_n$, see Fig.~\ref{fig:porosity}(b).

The results are dramatic in that they show very little variation in the global brick strength $\sigma$, even with upto $\pm 15\%$ variation in local stiffness. The result appears counterintuitive and offers the following practical recommendation. In order to increase the strength of individual bricks, it is perhaps more useful to reduce the global porosity instead of enhancing the stiffness of the local bonds. This has practical implications for the brick manufacturing process as we discuss in Sec.~\ref{sec:discussion}.

\subsection{Node coordination and fracture initiation}

\begin{figure}[h!]
  \centering
\begin{subfigure}[b]{0.7\columnwidth}
  \includegraphics[width=\linewidth]{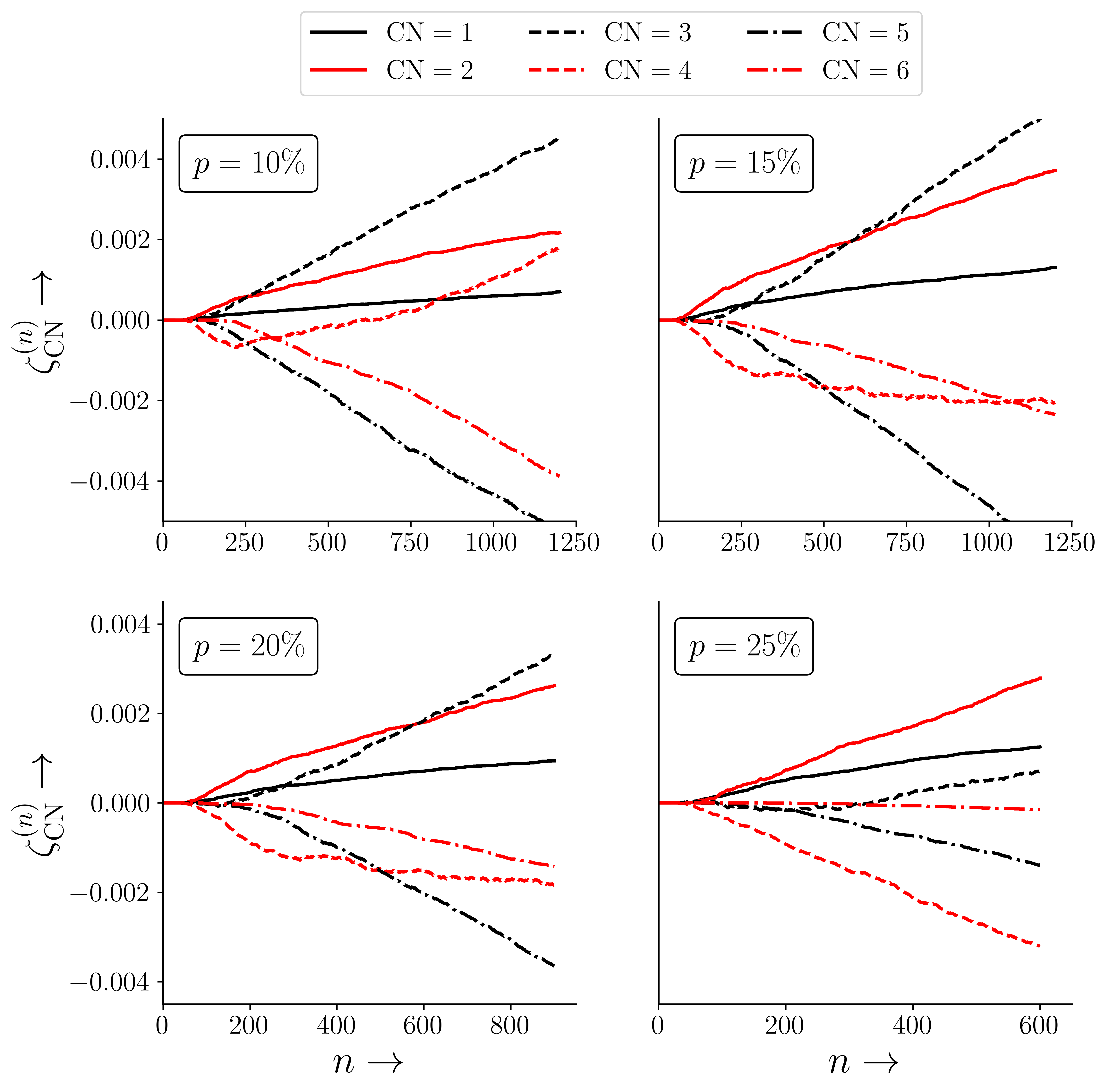}
  \caption{Variation in coordination number with increased loading}
\end{subfigure}
\begin{subfigure}[b]{0.49\columnwidth}
  \includegraphics[width=\linewidth]{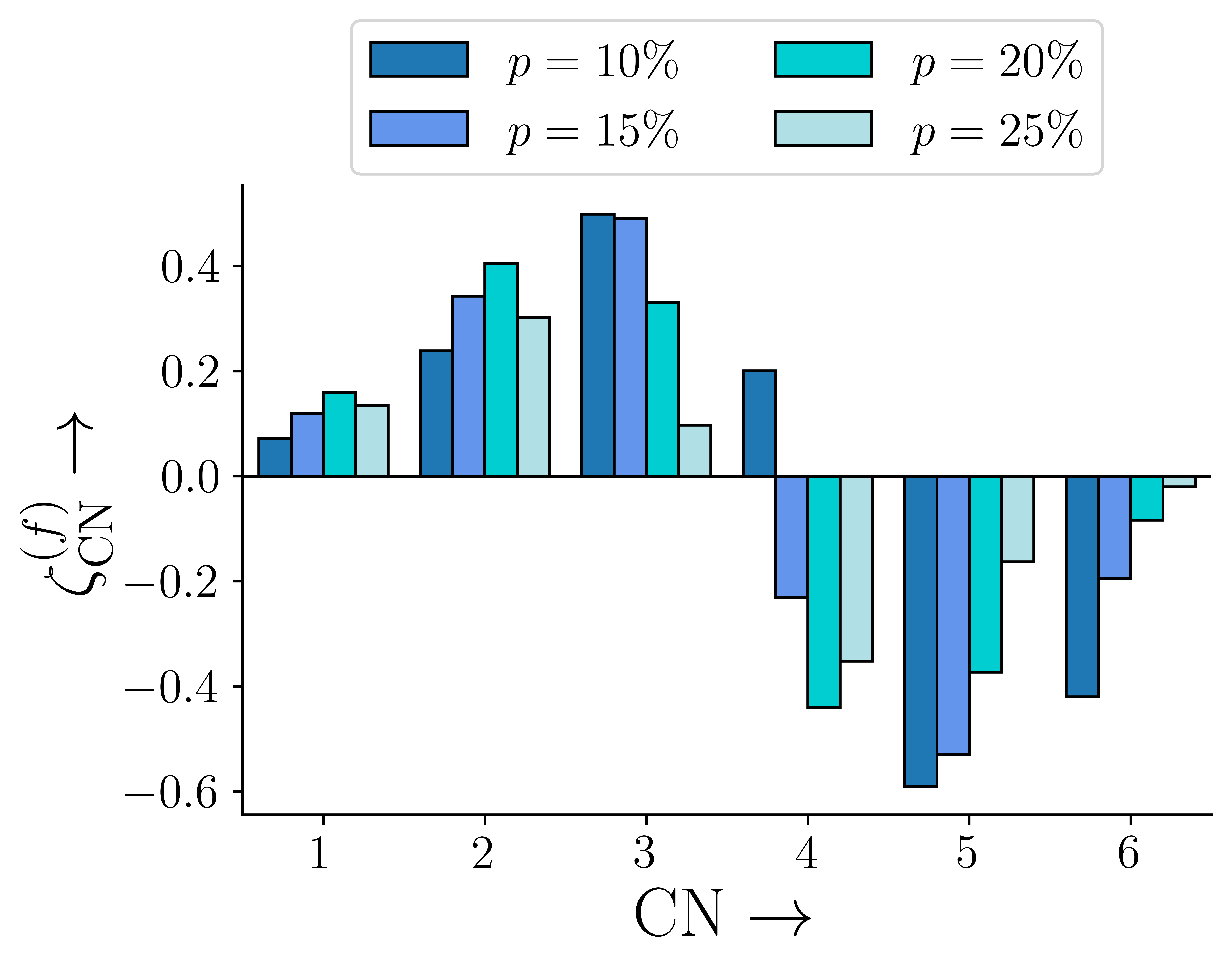}
  \caption{Statistics of final coordination $\zeta_{\text{CN}}^{(f)}$.}
\end{subfigure}
\caption{Change in coordination number with porosity. Simulation parameters: Lattice size = $300 \times 300$, Number of realization = 3. All other parameters are same.}
\label{fig:coordination}
\end{figure}

We now discuss the statistics of the fracture process within the lattice model, by taking recourse to local geometric computations. Firstly, we define the coordination number (CN) of a node in the usual manner: it is the number of nearest neighbours with which the node has an unbroken bond. Given the geometry of the lattice, CN varies from 1-6 for any node in the lattice; larger the value, the more rigid the entire network will be. The deformation response is reflected in the fractional change in coordination number. We define a quantity $\zeta_{\text{CN}}^{(n)}$ that is evaluated at each simulation iteration $n$ for the entire lattice
\begin{align*}
\zeta_{\text{CN}}^{(n)} = \dfrac{\text{CN}^{(n)}-\text{CN}^{(0)}}{N_t}
\end{align*}
where $\text{CN}^{(0)}$ and $\text{CN}^{(n)}$ are the number of nodes with coordination number (CN) at initial and $n^{th}$ iteration, respectively. $N_t$ is the total number of nodes in the network. The variation of this quantity ($\zeta_{\text{CN}}^{(n)}$) during the simulation process for different porosity is shown in Fig.~\ref{fig:coordination}(a).

The coordination number at a node effectively reveals information about the local pore morphology and load-bearing capability at each point in the material. A reduction in $\zeta_{\text{CN}}^{(n)}$ indicates a potentially dominant CN-node structure in the lattice from where fracture occurs. It is clear from Fig.~\ref{fig:coordination}(a) that $\zeta_4^{(n)}$ decreases first at the beginning of the fracture process, irrespective of the porosity. This means that for any given network, fracture propagation likely occurs from 4-coordinated nodes and these may hence be \emph{a priori} identified as potential fracture initiation spots. Likewise, the dominance of breakdown of 5-node structures decreases with higher porosity values.

Additional information may be gleaned by comparing the reduction of $\zeta_{4}^{(n)}$ and increase of $\zeta_{3}^{(n)}$. For $p = 25\%$, $\zeta_{4}^{(n)}$ decreases very rapidly, so that it should decompose to 3, 2, or 1-node structures. Consequently, $\zeta_{3}^{(n)}$, $\zeta_{2}^{(n)}$ or $\zeta_{1}^{(n)}$ should increase continuously with $n$. The relatively larger increase of $\zeta_{2}^{(n)}$ vis-\'a-vis $\zeta_{3}^{(n)}$ is only possible if 4-node decompose to decompose 3-node and then to 2-node. This situation is a signature of either local crack branching or multiple crack nucleation and growth. We have already seen that both of these phenomena predominate as the porosity is increased, \emph{cf.} Fig.~\ref{fig:fractureSequence}.

\subsection{Crack growth, branching and merging}

\begin{figure}[ht!]
	\centering
	\begin{subfigure}[b]{0.49\columnwidth}
		\includegraphics[width=\linewidth]{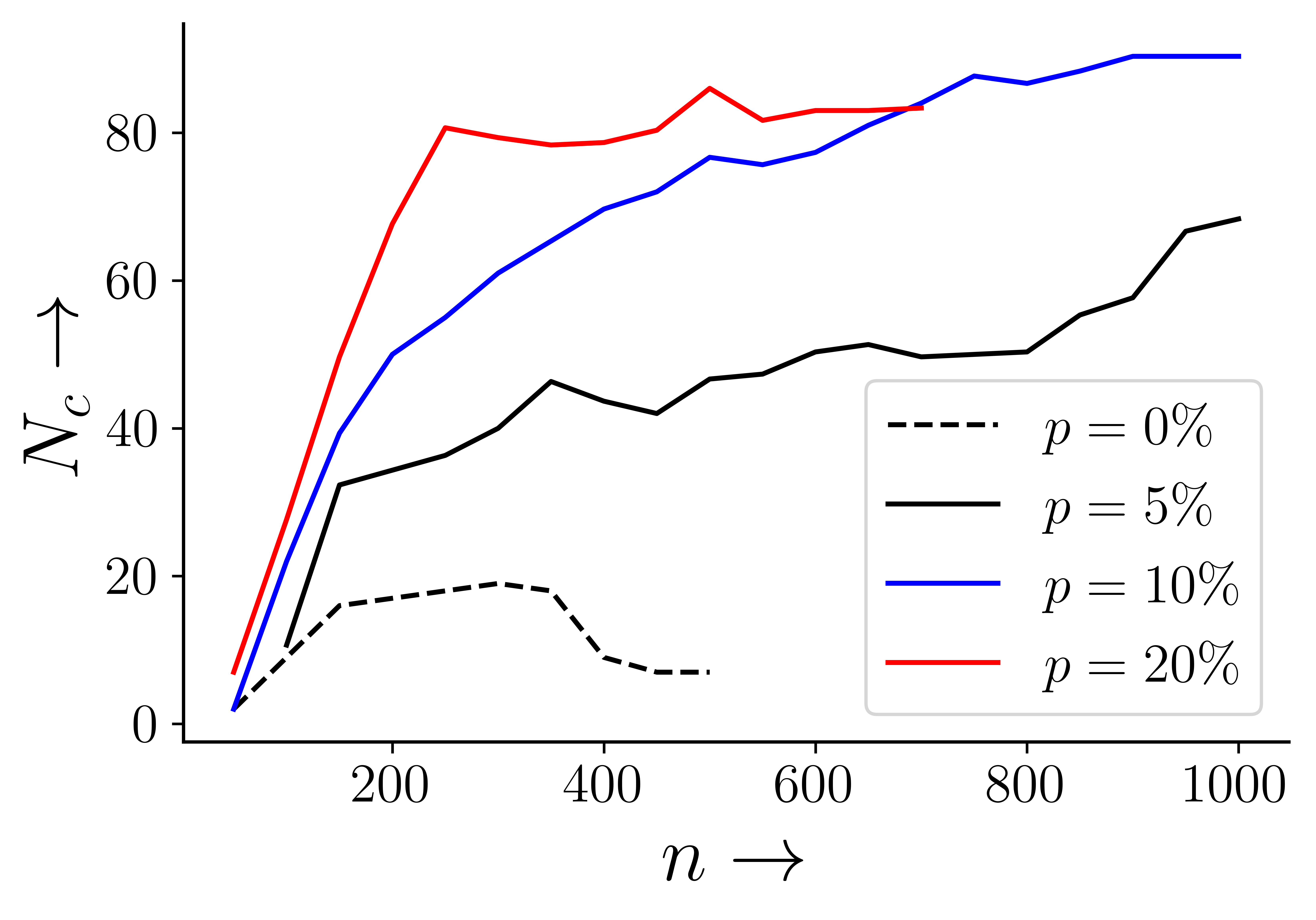}
		\caption{}
	\end{subfigure}
	\begin{subfigure}[b]{0.49\columnwidth}
		\includegraphics[width=\linewidth]{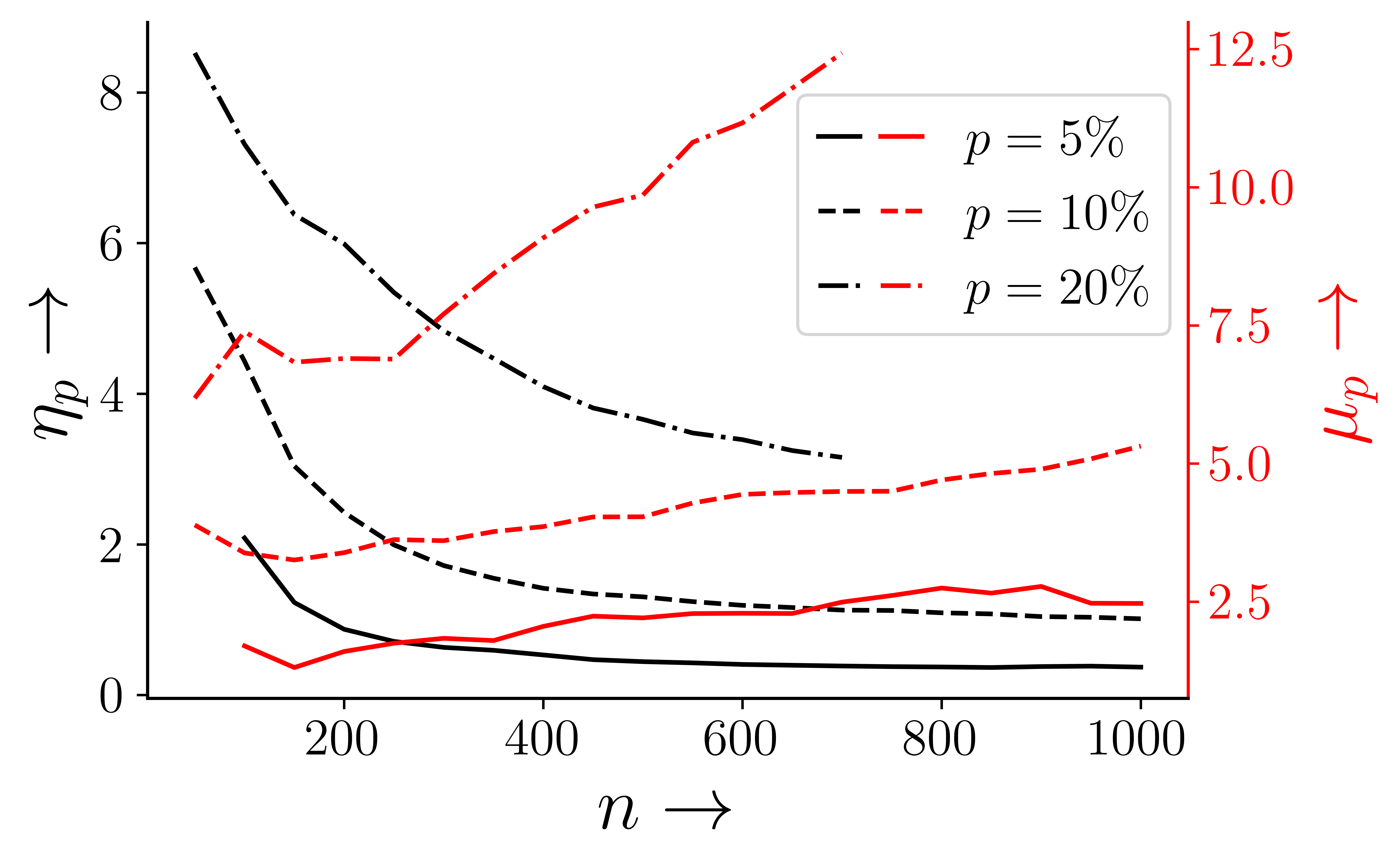}
		\caption{}
	\end{subfigure}
	\caption{Simulated fracture analysis for different porosities. Simulation parameters: Lattice size = $200 \times 200$. All other parameters are same.}
	\label{fig:crackNumbers}
\end{figure}

To further analyze the fracture mechanism, especially the evolution, branching and merging of single cracks, we evaluated the parameter called crack number ($N_c$), defined as the number of primary (main branch) cracks in the lattice network. As defined, $N_c$ is indicative of either new crack formation or of merging of existing cracks. The latter results in a single primary crack. Constant value of $N_c$ indicates propagation of multiple cracks or nucleation and merger occurring simultaneously. As a baseline measure, we first quantify $N_c$ for a homogeneous lattice $p=0\%$, see dashed line in Fig.~\ref{fig:crackNumbers}(a). In this case, we explicitly introduced an inclined central crack for initiation fracture. In the initial stages, a few micro-cracks were first observed near the tip of the main crack, which eventually merged into a single crack quickly reuslting in complete fracture. In contrast, for a porous lattice $p\neq 0$, crack nucleation, propagation, and merging are continuous processes that occur simultaneously, see Fig.~\ref{fig:crackNumbers}(a). Also, it is observed that the fracture process is delayed due to the presence of the pores; this agrees with the known fact that pores can act as crack deflectors.

To account for the number of pores entrapped along the crack path, we have estimated a quantity called pore to crack ratio ($\eta_p$) for a porosity $p$, defined as the sum of the all pore sizes ($d$) that coalesce with primary cracks path to the total length of all the primary cracks, and can be expressed as
\begin{equation}
\eta_p = \frac{\sum_j d^{(j)}}{\sum_i^{N_c}a_i}
\end{equation}
where $d^{(j)}$ is the $j^{th}$ pore connected with the primary crack path, and $a_i$ is the length of the $i$th primary crack path. This ratio $\eta_p$ corresponds to the pore density of the path traced by the primary crack. The $\eta_p$ decreases rapidly initially and saturates to some constant value for all porosity, see Figure ~\ref{fig:crackNumbers}(b). This value is higher for higher porosity, indicating more pores along the crack path, hence more chances of propagating crack to be deflected, justifying the horizontal cracks observed for higher porosity.

%A second quantity called the pore to crack ratio ($\eta_p$) is also evaluated. $\eta_p$ is defined as the sum of the all pore sizes ($d$) that coalesce with the path of primary cracks, divided by the total length of all primary cracks:
%\begin{align}
%\eta_p = \frac{\sum_j d^{(j)}}{\sum_i^{N_c}a_i}
%\end{align}
%where $d^{(j)}$ is the $j$th pore connected with a primary crack path and $a_i$ is the length of the $i$th primary crack. 

A third statistical quantity that evaluates the number of active pores is termed the mean active pore size ($\mu_p$), and defined as the sum of all pore sizes that coalesce with the crack, divided by the total number of cracks. An increase in $\mu_p$ implies that cracks propagate between pores and connect them. Among all the simulation runs, we observed that $\mu_p$ increased as a function of $n$ all values of porosity, but the rate of increase is higher for $p = 20\%$. We are presently exploring the consequences of both $\eta_p$ and $\mu_p$ variations with $n$ in more detail. It is likely that they encode additional details of the crack paths and their interaction with pores within the lattice network.

%We also have estimated a quantity called pore size ($d$) to crack size($a_c$) ratio is defined as the sum of the sizes of all the pores that coalesce with cracks to the total length of all the cracks. Physically, it indicates the pore density of the path traced by crack. This quantity decreases rapidly initially and saturates to some constant value for all porosity, see Figure ~\ref{fig: figure7}(b). This value is higher for higher porosity, indicating more pores along the crack path, hence more chances of propagating crack to be deflected, justifying the horizontal cracks observed for higher porosity.

%Another useful measure is termed 'mean active pore size' (MAPS), defined as the sum of all pore sizes that coalesce with the crack, divided by the total number of cracks. 
%$MAPS = \dfrac{\text{Mean active pore}}{\text{Total active pore}}*{\text{Number of pore}}$
%The increase or decrease in MAPS respectively reflects whether the new cracks initiated from the pores or not. The increase in MAPS, is due to either by initiation of new crack or by presence of more pores in the same crack. Hence, for porous material MAPS is higer.

\section{Discussion and Summary}
\label{sec:discussion}

A comprehensive understanding of the fracture behaviour and deformation response of consolidated lunar regolith is important if they are to be exploited for critical load bearing applications. Given the inherently stochastic nature of the consolidation process, and the importance of porosity in the final brick structures, predicting fracture responses have remained important challenges. In this work, we have demonstrated how a numerical framework based on a lattice network model can capture several of the critical features of the fracture process of consolidated space bricks, and provide microscopic insight into their deformation response that is otherwise inaccessible by conventional macroscale mechanical testing methods. We now discuss some of the implications of our results within the broader context of consolidated regoliths and suggest possible adaptations of our work to understand other similar systems in more detail, for instance, in Refs.~\cite{lauermannova2021regolith, oh2021impact, warren2022effect} or recent reviews on this topic \cite{toklu2022lunar, farries2021sintered}. 

Firstly, the network model may be thought of as an analogue of conventional finite element discretization schemes, albeit accounting for local pore distributions. In fact, the final equilibrium configuration is determined by solving an analogous linear system (Eq.~\ref{eqn:governing}) in our case, with an additional local rotational degree of freedom that allows for change in local pore shapes. 

Secondly, our results have shown that the lattice network model, when suitably calibrated against experimental observations for the two parameters $k_n$ and $\sigma_c$, is capable of not only capturing stress strain curves for various consolidation conditions (changing global porosity, local bond stiffness variations) but also provides a detailed microscopic picture of how cracks nucleate, potentially merge, and propagate, leading to final failure. Furthermore, the model can then be used to predict strength variations under these conditions, \emph{cf.} Fig.~\ref{fig:porosity}. This is important because global porosity is not an easily controllable parameter in experiments and hence systematically investigating its effect on sample strength is beyond the capability of current experimental investigations.   

Finally, statistics of the network reveal signatures of the occurrence of crack branching as well as multiple cracks within the sample. It is expected that a simple extension of our 2D model to 3D will then enable identification and delineation of through-thickness micro-cracks that do not contribute to sample failure and in-plane macro cracks that do. Such a delineation is presently difficult to realized experimentally unless recourse is taken to sophisticated yet size-limited techniques such as \emph{in situ} X-ray micro-CT testing.

We believe that an approach similar to the one presented in this manuscript will be very useful in designing specific consolidated structures for specific strengthening applications such as, for instance, crack deflection and arrest. We are currently investigating some of these ideas using detailed simulations and coupled experimental analyses and hope to present our results in a forthcoming manuscript.

\section*{Acknowledgements}
The authors acknowledge Dr. Rashmi Dikshit for help with MICP and bioconsolidation experiments.

\clearpage
\bibliography{bibfile}
\bibliographystyle{vancouver}

\end{document}